\documentclass[journal]{IEEEtran}

\usepackage{url}
\usepackage{breakurl}
\usepackage[breaklinks]{hyperref}
\usepackage{amsmath,amssymb,amsfonts}
\usepackage{graphicx}
\usepackage{textcomp}
\usepackage{xcolor}

\def\BibTeX{{\rm B\kern-.05em{\sc i\kern-.025em b}\kern-.08em
    T\kern-.1667em\lower.7ex\hbox{E}\kern-.125emX}}
\usepackage[linesnumbered,ruled,vlined]{algorithm2e}
\usepackage{listings}
\usepackage[noend]{algpseudocode}
\SetCommentSty{mycommfont}
\SetKwInput{KwInput}{Input}                
\SetKwInput{KwOutput}{Output}              
\usepackage{subcaption}
\DeclareMathOperator{\sinc}{sinc}
\usepackage{balance}
\usepackage{cite}
\usepackage{soul}
\usepackage{acro} 
\DeclareAcronym{RIS}{
  short = RIS ,
  long  = reconfigurable intelligent surfaces ,
  class = abbrev
}
\DeclareAcronym{6G}{
  short = 6G,
  long  = 6th generation ,
  class = abbrev
}
\DeclareAcronym{5G}{
  short = 5G,
  long  = 5th generation ,
  class = abbrev
}
\DeclareAcronym{EM}{
  short = EM ,
  long  = electromagnetic,
  class = abbrev
}
\DeclareAcronym{BS}{
  short = BS,
  long  = base station ,
  class = abbrev
}
\DeclareAcronym{UE}{
  short = UE ,
  long  = user equipment ,
  class = abbrev
}
\DeclareAcronym{MISO}{
  short = MISO,
  long  = multiple-input single-output ,
  class = abbrev
}
\DeclareAcronym{MMSE}{
  short = MMSE ,
  long  = minimum mean squared error ,
  class = abbrev
}
\DeclareAcronym{DFT}{
  short = DFT,
  long  = discrete Fourier transform ,
  class = abbrev
}
\DeclareAcronym{THz}{
  short = THz,
  long  = Terahertz ,
  class = abbrev
}
\DeclareAcronym{IoT}{
  short = IoT,
  long  = internet of things ,
  class = abbrev
}
\DeclareAcronym{MSE}{
  short = MSE,
  long  = mean square error ,
  class = abbrev
}
\DeclareAcronym{CSI}{
  short = CSI ,
  long  = channel state information ,
  class = abbrev
}
\DeclareAcronym{MIMO}{
  short = MIMO,
  long  = multiple-input multiple-output ,
  class = abbrev
}
\DeclareAcronym{UPA}{
  short = UPA,
  long  = uniform planner array ,
  class = abbrev
}
\DeclareAcronym{RF}{
  short = RF,
  long  = radio-frequency ,
  class = abbrev
}
\DeclareAcronym{mmWave}{
  short = mmWave,
  long  = millimeter-wave ,
  class = abbrev
}
\DeclareAcronym{AoA}{
  short = AoA ,
  long  = angle of arrival ,
  class = abbrev
}
\DeclareAcronym{AoD}{
  short = AoD,
  long  = angle of departure ,
  class = abbrev
}
\DeclareAcronym{EKF}{
  short = EKF,
  long  = extended Kalman filter ,
  class = abbrev
}
\DeclareAcronym{UKF}{
  short = UKF,
  long  = unscented Kalman filter ,
  class = abbrev
}
\DeclareAcronym{LMS}{
  short = LMS,
  long  = least mean square ,
  class = abbrev
}
\DeclareAcronym{BiLMS}{
  short = BiLMS,
  long  = bi-directional LMS ,
  class = abbrev
}
\DeclareAcronym{SNR}{
  short = SNR,
  long  = signal-to-noise ratio ,
  class = abbrev
}
\DeclareAcronym{LoS}{
  short = LoS,
  long  = line-of-sight ,
  class = abbrev
}
\DeclareAcronym{NLoS}{
  short = NLoS,
  long  = non-line-of-sight ,
  class = abbrev
}
\DeclareAcronym{TDD}{
  short = TDD,
  long  = time-division duplexing ,
  class = abbrev
}
\DeclareAcronym{NMSE}{
  short = NMSE,
  long  = normalized mean square error ,
  class = abbrev
}
\DeclareAcronym{SDR}{
  short = SDR,
  long  = semidefinite relaxation ,
  class = abbrev
}
\DeclareAcronym{QoS}{
  short = QoS,
  long  = quality of service ,
  class = abbrev
}
\DeclareAcronym{NOMA}{
  short = NOMA,
  long  = non-orthogonal multiple access ,
  class = abbrev
}
\DeclareAcronym{OMA}{
  short = OMA,
  long  = orthogonal multiple access ,
  class = abbrev
}
\DeclareAcronym{NU}{
  short = NU,
  long  = near user ,
  class = abbrev
}
\DeclareAcronym{FU}{
  short = FU,
  long  = far user ,
  class = abbrev
}
\DeclareAcronym{SIC}{
  short = SIC,
  long  = successive interference cancellation ,
  class = abbrev
}
\DeclareAcronym{PLS}{
  short = PLS,
  long  = physical layer security ,
  class = abbrev
}
\DeclareAcronym{MRT}{
  short = MRT,
  long  = maximum ratio transmission ,
  class = abbrev
}
\DeclareAcronym{AWGN}{
  short = AWGN,
  long  = additive white Gaussian noise,
  class = abbrev
}
\DeclareAcronym{SINR}{
  short = SINR,
  long  = signal-to-interference-plus-noise ratio ,
  class = abbrev
}
\DeclareAcronym{BPSK}{
  short = BPSK,
  long  = binary phase shift keying ,
  class = abbrev
}
\DeclareAcronym{QPSK}{
  short = QPSK,
  long  = quadrature phase shift keying ,
  class = abbrev
}
\DeclareAcronym{SVD}{
  short = SVD,
  long  = singular value decomposition ,
  class = abbrev
}
\DeclareAcronym{PDF}{
  short = PDF,
  long  = probability density function ,
  class = abbrev
}
\DeclareAcronym{SER}{
  short = SER,
  long  = symbol error rate ,
  class = abbrev
}
\DeclareAcronym{MGF}{
  short = MGF,
  long  = moment generating function ,
  class = abbrev
}
\DeclareAcronym{2D}{
  short = 2D,
  long  = two-dimensional ,
  class = abbrev
}
\DeclareAcronym{3D}{
  short = 3D,
  long  = three-dimensional ,
  class = abbrev
}
\DeclareAcronym{CLT}{
  short = CLT,
  long  = central limit theorem ,
  class = abbrev
}
\DeclareAcronym{QAM}{
  short = QAM,
  long  = quadrature amplitude modulation ,
  class = abbrev
}
\DeclareAcronym{SISO}{
  short = SISO,
  long  = single-input single-output ,
  class = abbrev
}
\DeclareAcronym{CE}{
  short = CE,
  long  = channel estimation ,
  class = abbrev
}
\DeclareAcronym{LAA}{
  short = LAA,
  long  = lens antenna array ,
  class = abbrev
}
\DeclareAcronym{ULA}{
  short = ULA,
  long  = uniform linear array ,
  class = abbrev
}
\DeclareAcronym{UT}{
  short = UT,
  long  = unscented transformation ,
  class = abbrev
}
\DeclareAcronym{UL}{
  short = UL,
  long  = uplink ,
  class = abbrev
}
\DeclareAcronym{DL}{
  short = DL,
  long  = downlink ,
  class = abbrev
}

\definecolor{GreenForest}{rgb}{0.09, 0.45, 0.27}

\begin{document}

\title{Joint Tracking of Multiple Beams\\in Beamspace MIMO Systems}

\author{
     \IEEEauthorblockN{Liza Afeef,} 
    \IEEEauthorblockN{Murat Karabacak,}
	\IEEEmembership{Member, IEEE,} 
	\IEEEauthorblockN{and}
	\IEEEauthorblockN{H\"{u}seyin Arslan,}
	\IEEEmembership{Fellow, IEEE}

\thanks{L. Afeef and H. Arslan are with the Department of Electrical and Electronics Engineering, Istanbul Medipol University, Istanbul, 34810, Turkey (e-mail: liza.shehab@std.medipol.edu.tr;  huseyinarslan@medipol.edu.tr).}
\thanks{M. Karabacak and H. Arslan are with Department of Electrical Engineering,
	University of South Florida, Tampa, FL, 33620, USA (e-mail: murat@usf.edu; arslan@usf.edu).}

\thanks{This work has been submitted to the IEEE for possible publication. Copyright may be transferred without notice, after which this version may no longer be accessible.}
}

\maketitle

\begin{abstract}
In millimeter-wave (mmWave) systems, beamforming is needed to overcome harsh channel environments. As a promising beamforming solution, lens antenna array (LAA) implementation can provide a cost-effective solution without notable performance degradation compared to its counterpart. However, an appropriate beam selection is a challenge since it requires efficient channel estimation via an extensive beam training process for perfect beam alignment. In this paper, we propose a high mobility beam and channel tracking algorithm based on the unscented Kalman filter (UKF) to address this challenge, where the channel changes can be monitored over a certain time. The proposed algorithm tracks the channel changes after establishing a connection with an appropriate beam. The algorithm is first introduced in a multi-user beamspace multiple-input multiple-output (MIMO) system with LAA where a single beam is tracked at the user side at downlink transmission. Then, it is employed for multi-beam joint-tracking at the base station side in the uplink transmission. The analysis indicates that under different channel variation conditions, the proposed algorithm outmatches the popular extended Kalman filter (EKF) in both single-beam and multi-beam tracking systems. While it is common to individually track the beams in a multi-beam system scenario, the proposed joint tracking approach can provide around 62\% performance enhancement compared to individual beam tracking using the conventional EKF method.
\end{abstract}

\IEEEpeerreviewmaketitle

\section{Introduction}

\IEEEPARstart{M}ILLIMETER-WAVE (mmWave) communication is considered as a promising technology to support the envisioned high data rate in next-generation wireless networks \cite{pi2011introduction}. However, mmWave signals suffer from a severe path loss problem due to harsh propagation conditions including blockage at high frequencies \cite{heath2016overview, dougan2018optimization}. Therefore, \ac{MIMO} beamforming implementation is merged in mmWave communications to achieve highly directional beams and mitigate such path loss effects. Apart from combating the path loss problem, beamforming also reduces interference, boosts capacity, enhances security \cite{pekoz2020reducing}, and offers better coverage at the cell edge \cite{ahmed2018survey, molisch2017hybrid}.

Despite the traditional beamforming approach of phase shifters for each antenna aperture which causes high power consumption, the studies in \cite{brady2013beamspace, zeng2016millimeter, al2011wideband} uses a lens on top of the antenna array, and with switches in the place of phase shifters. Such implementation of a lens into antenna array, referred to as \ac{LAA}, exhibits some distinctive properties \cite{cho2018rf}; 1) It focuses signal power at the front-end to achieve high directivity, 2) it concentrates signal power directed to a sub-region of the antenna array, and 3) it replaces the phase shifters with switches which reduce the cost. Owing to these properties and advantages, lens antenna systems are highly considered to be an effective solution for mmWave communications in terms of cost and performance \cite{LensMagazine}. \ac{LAA} systems can also offer high gain and relatively low sidelobes in different directions without any significant performance loss \cite{zeng2014electromagnetic, kwon2016rf}. Designing an \ac{LAA} system whose aperture phase distribution equalized in a scanning plane is a straightforward procedure \cite{lau2013electromagnetic}. Additionally, an increased level of channel sparsity in mmWave \ac{LAA} systems, makes it possible to improve the channel estimation using dictionary-based sparse estimators \cite{nazzal2019dictionary}.
Besides, By employing the \ac{LAA}, the spatial channel representation can be converted to the beamspace channel model \cite{zeng2014electromagnetic}.

The sparse nature of the beamspace \ac{MIMO} channel allows selecting a small number of beams and thus reduces the number of \ac{RF} chains in the system. However, an accurate \ac{CSI} is needed in beamspace \ac{MIMO} systems \cite{wei2020deep} which require to have frequent estimating for the channel leading to a huge overhead and large loss of throughput \cite{yang2017channel,li2017millimeter}. Such often channel estimation can be avoided using tracking algorithms to track the channel parameters, i.e. channel coefficient, \ac{AoA}, and \ac{AoD}. 
The beam tracking algorithms are significantly fast, reliable, and robust which allow efficient data transfer between transmitters and receivers in mmWave communications.
In high dynamic communication networks, channel tracking can overcome the performance degradation of physical layer authentication \cite{bai2020physical}.

\subsection{Prior works}
Several works are proposed regarding beam tracking techniques for mmWave communication. The first work in this direction is presented in \cite{zhang2016tracking}, where an analog beamforming strategy is selected and an \ac{EKF} based tracking algorithm for a sudden change detection method is proposed to track \ac{AoA}/\ac{AoD} while assuming constant channel coefficient in a mmWave system. This filter uses Jacobian matrices to transform the non-linear system into linear approximations around the current state. The results show that the use of the \ac{EKF} algorithm causes a gracious decaying in the system performance with the acquisition error while requiring a low \ac{SNR} and low pilot overhead. The method has difficulties to track in a fast-changing channel environment since it requires pre-requisites for a full scan that causes long time measurement. To decrease the measurement time and provide a more suitable tracking algorithm, the authors in \cite{va2016beam} proposed an alternative solution that requires only a single measurement with \ac{EKF} estimation and a beam switching design. As an extension for the work in \cite{va2016beam}, the authors in \cite{jayaprakasam2017robust} proposed a joint minimum \ac{MSE} beamforming with the help of \ac{EKF} tracking strategy. Using same filter, \cite{shaham2019fast} proposes a beam tracking model for motion tracking (position, velocity, and channel coefficient) in mmWave vehicular communication system. The main different of this model is shown in its state variables where approximate linear motion equations are derived from the beam angles to avoid the nonlinearity of using angles in the state variables which reduces the complexity in calculating the Jacobians matrix.

However, the above-discussed techniques are limited to the scenarios where only a single beam is considered or multiple beams with uncorrelated paths. In \cite{fan2018stochastic}, a Markov jump linear system (MJLS) and an optimal linear filter are designed to track the dynamics of the channel with two beams considering the correlation between them. This system iteratively tracks the \ac{AoA} of the incoming beams only, taking into account the channel gain correlation between different paths. However, the computational complexity of this method increases exponentially with the number of target beams. 
In a faster angle variation environment, \cite{lim2019beam} proposes a tracking algorithm based auxiliary particle filter (APF) that displays optimal performance using 32 antennas. Although APF shows improved performance, this approach requires a high processing time compared to \ac{EKF} approach. 
In \cite{zhu2018high}, angle tracking strategies for wideband mmWave systems are proposed where pairs of auxiliary beams are designed as the tracking beams to capture the angle variations, toward which the steering directions of the data beams are adjusted. The proposed methods are independent from a particular angle variation. However, their analysis show that the method is sensitive to radiation pattern impairments.

In mmWave beamspace \ac{MIMO} system with multiple \ac{UE}s, only the work in \cite{dai2016priori} proposes a channel tracking algorithm by exploiting the temporal correlation of the time-varying channels to track the \ac{LoS} path of the channel. Considering a motion model for the \ac{UE}s, a temporal variation law of the \ac{AoA} and \ac{AoD} of the \ac{LoS} path is excavated and tracked based on the sparse structure of beamspace \ac{MIMO} system. However, a large number of pilots are implemented to employ the tracking with the presented method. 

\subsection{Our contribution}
Despite of the benefit of utilizing LAA in MIMO systems, the literature lack the sufficient analysis of tracking approaches in these systems. Although few researches paid attention to various types of channel tracking filters \cite{va2016beam,jayaprakasam2017robust,lim2019beam}, none of them spots the light on tracking multi-beam jointly in a multi-beam beamspace \ac{MIMO} systems to provide multiple users support at the same time. In this paper, we introduce multiple user support while implementing multi-beam tracking with a reduced complexity algorithm based on \ac{UKF}. The contributions of this paper are summarized as follows:

\begin{itemize}    
    \item  \ac{LAA} concept is considered as a practical solution for the future mmWave communication systems since it can provide less hardware complexity and high antenna gains. However, such beamforming systems still in need of beam tracking algorithms for efficient usage of the beams. Thus, in this paper for the first time, a beam tracking algorithm is implemented in an \ac{LAA} system to the best of authors' knowledge.
    \item Due to the ability of the \ac{UKF} algorithm to adapt in a high dynamic state estimation, for the first time, \ac{UKF} is adapted to track channel parameters (i.e. \ac{AoA}, \ac{AoD}, and directional channel coefficients) of a multi-user beamspace \ac{MIMO} communication system, where the algorithm parameters and steps are optimized to properly work on this system.
    \item The tracking algorithm is adapted to beamspace \ac{MIMO} communication by optimizing the sigma points spreading parameters of the \ac{UKF} to provide optimized solution for \ac{DL} and \ac{UL} transmission scenarios where single-beam and multi-beam tracking are applied, respectively.
    \item The performance of the \ac{UKF} algorithm-based tracking is compared with the \ac{EKF} algorithm. Evaluation results indicate that the proposed \ac{UKF} algorithm outperforms the conventional \ac{EKF} algorithm while having the same complexity. 

\end{itemize}

\subsection{Organization and notation}
This paper is organized as follows. In Section \ref{sec:system-model}, the system model is presented for \ac{DL} and \ac{UL} transmission where the beamspace channel transformation is introduced also. The frame structure and evolution models of the whole system along with the proposed strategy of the \ac{UKF} approach are provided in Section \ref{sec:tracking-algorithm}. In Section \ref{sec:numerical-results}, simulation results for each approach are carried out to evaluate the performance of the algorithm compared to those by conventional \ac{EKF}. Finally, conclusions and future vision are given in Section \ref{sec:conclusion}.

\emph{Notation}: Matrices are denoted by bold uppercase letters (e.g. $\boldsymbol{A}$), and vectors are denoted by bold lowercase letters (e.g. $\boldsymbol{a}$). $\boldsymbol{A}^T$ and $\boldsymbol{I}_{Z}$ denote the Hermitian (conjugate transpose) of matrix $\boldsymbol{A}$ and $Z\times Z$ identity matrix, respectively. $\sinc()$ is the “sinc” function defined as $\sinc(x) = \sin(\pi~ x)/(\pi~ x)$, and for a real number $A$, $\lfloor \cdot \rfloor$ denotes a floor operation. Furthermore, $E\{\cdot\}$ denotes the expectation operator.
\begin{figure}
\centering
\includegraphics[scale=0.23]{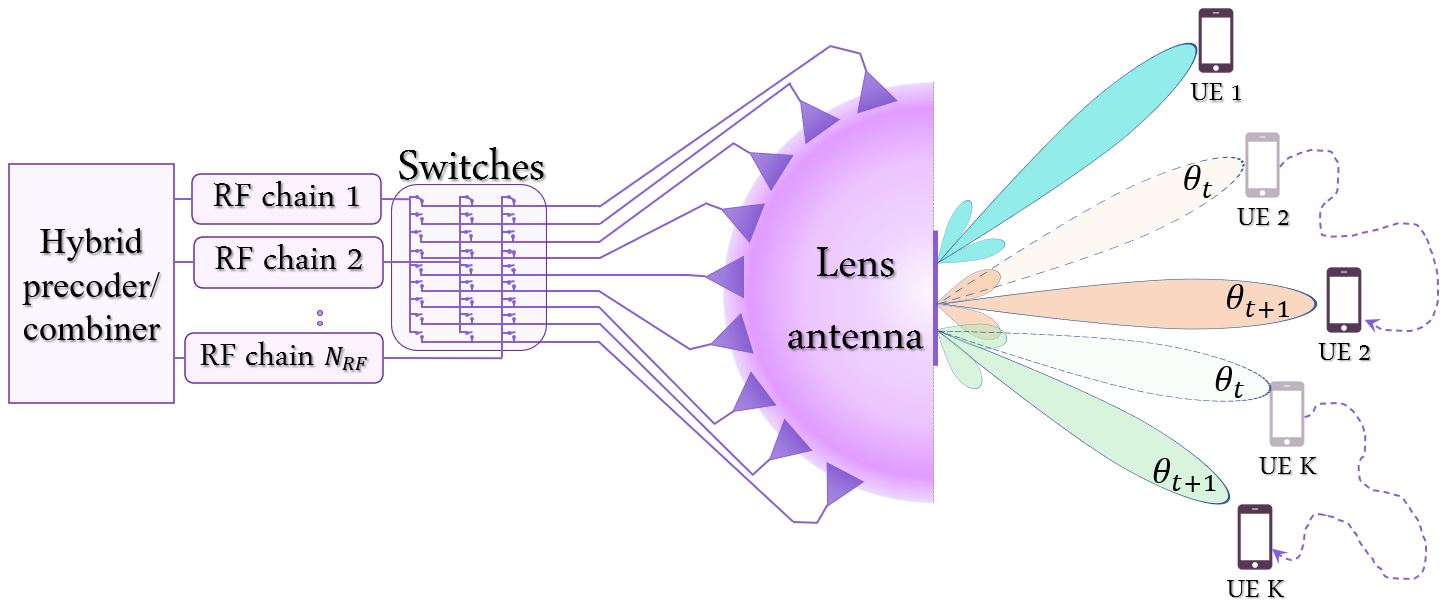}
\caption{System model for the beamspace mmWave \ac{MIMO} with \ac{LAA}.}
\label{fig:systemmodelall}
\end{figure}
\section{System Model} \label{sec:system-model}

A typical mmWave beamspace \ac{MIMO} system is considered where the \ac{BS} employs $N_{\operatorname{BS}}$ antennas and $N_{\operatorname{RF}}$ \ac{RF} chains to serve $K$ \ac{UE}s with single \ac{RF} chain and $N_{K}$ antennas. 
Assuming that the system has \ac{2D} motion model where the azimuth angle $\theta$ is needed here only, but the extension to elevation and azimuth is possible \cite{zhu2018high}. 

The received noisy signal for user $k$ at time slot $t$ is given as
\begin{equation}
\label{equ:received-signal-downlink}
\begin{aligned}
  {y}_{k,t} = \boldsymbol{w}^H_k \mathbf{H}_{k,t} \mathbf{p}_k {s}_k +  \underbrace{\boldsymbol{w}^H_k  \mathbf{H}_{k,t} \sum_{i=1, i\neq k}^{K} \mathbf{p}_i {s}_i} + \boldsymbol{w}^H_k {v}_{t} ,\\
  ~\operatorname{Inter-users~interference} ~~~~
\end{aligned}
\end{equation}
where $\boldsymbol{w}_k$ is $\mathbb{C}^{N_k \times 1}$ combiner vector for the $k$-th user, $\mathbf{H}_t = \left[\mathbf{H}_{1,t}, \mathbf{H}_{2,t}, \dots , \mathbf{H}_{K,t}\right]$ is the channel matrix, $\mathbf{H}_{k,t}$ is the channel matrix between the BS and the $k$-th user, $\mathbf{P} = [\mathbf{p}_1, \mathbf{p}_2, \dots , \mathbf{p}_K]$ is the hybrid beamformer matrix, $\mathbf{s}=[s_1, s_2, \dots, s_K]$ are the transmitted symbol for all users with normalized power, and ${v} \sim \mathbb{C}\mathcal{N}(\mathbf{0},\sigma_v^2)$ is the additive white Gaussian noise.

In general, it is considered that the channel between \ac{BS} and each \ac{UE} $\mathbf{H}_k$ follows a geometrical narrowband slow time-varying channel model \cite{heath2016overview}, given as
\begin{equation} \label{equ:spatial_channel}
\begin{aligned}
  \mathbf{H}_{k,t}  = \sqrt{\frac{N_{\operatorname{BS}}N_k}{L_k}} \sum_{l=1}^{L_k}\alpha_{k,l,t} ~\boldsymbol{a}(\theta_{k,A,l,t}) \boldsymbol{a}^H(\theta_{k,D,l,t}),
\end{aligned}
\end{equation}
where $L_k$ is the number of resolvable channel paths between the \ac{BS} and user $k$, $\alpha_{k,l}$ is the complex channel coefficient, and $\theta_{k,A,l}$ and $\theta_{k,D,l}$ are the \ac{AoA} and \ac{AoD} for path $l$ of user $k$. $\boldsymbol{a}$ represents the steering vector for \ac{ULA} antenna which is given as \cite{wang2017iterative} 
\begin{equation} \label{equ:steering}
  \boldsymbol{a} = \frac{1}{\sqrt{N}}  [ e^{-j\frac{2\pi}{\lambda}\frac{-(N-1)}{2}d~sin(\theta)}, \dots , e^{-j\frac{2\pi}{\lambda}\frac{(N-1)}{2}d~sin(\theta) }]^T,
\end{equation}
where $\lambda$ is the carrier wavelength and $d$ is the antenna spacing satisfying $d=\lambda/2$, $N \in \{ N_{\operatorname{BS}},N_K\}$, $\theta \in \{ \theta_A, \theta_D \}$.

In order to decrease the effect of the users interference during the tracking, we propose to track the beams at the \ac{BS} side at the \ac{UL} mode of the system where all users are transmitted to be received at the same time by the \ac{BS}. Considering all the assumptions, the received noisy vector $\mathbf{y}$ from all $K$ users at time slot $t$ becomes
\begin{equation}
\label{equ:received-signal-uplink}
  \mathbf{y}_{t} = \boldsymbol{W}^H \mathbf{H}_t \mathbf{P} \mathbf{D} \mathbf{s} + \boldsymbol{v}_{t} ,
\end{equation}
where $\mathbf{D}=diag(\frac{1}{\sqrt{\rho_1}},\cdots,\frac{1}{\sqrt{\rho_{K}}})$ describes the average path loss between the \ac{BS} and each \ac{UE}. 

\subsection{Beamspace channel transformation} \label{sec:beamspace-channel}
In order to transfer the conventional spatial channel to a beamspace one, a \ac{LAA} is carefully designed at the \ac{BS} and \ac{UE} sides. Therefore, the channel in  \eqref{equ:spatial_channel} is transformed into beamspace channel as \cite{brady2013beamspace}
\begin{equation}
\begin{aligned}
  \mathbf{H}_{b,t} = [\mathbf{H}_{b,1,t}, \mathbf{H}_{b,2,t}, \dots , \mathbf{H}_{b,K,t}] , \\
  = \mathbf{U} \mathbf{H}_t \mathbf{U}^H ,~~~~~~~~~~~~~~~~ 
\end{aligned}
\end{equation}
where $\mathbf{H}_{b,k,t}$ is the beamspace channel of the $k$-th \ac{UE} and $\mathbf{U} = \frac{1}{\sqrt{N}} \left[ \mathbf{u}(\psi_0),~\mathbf{u}(\psi_1),~\dots, ~\mathbf{u}(\psi_{N-1}) \right]^H$ is a unitary \ac{DFT} matrix that uses to transform the spatial channel into beamspace channel where $\mathbf{u}(\psi_\ell)$ represents the virtual steering vector at specific virtual angle $\psi_\ell$ where $\psi_\ell=\frac{1}{N}\left( \ell - \frac{N+1}{2} \right)$ and each element in this vector is given as $u_n(\psi_\ell) = e^{-j2\pi \psi_\ell (n-\frac{N-1}{2})}, ~~n \in \{1,2,...,N-1\}$.

In general, the beamspace channel between each \ac{UE} and the \ac{BS}, either in \ac{UL} or \ac{DL} transmission, can be written as 
\begin{equation}
    \mathbf{H}_{b,k,t} = \sqrt{\frac{N_t N_r}{L_k}} \sum_{l=1}^{L_k} \alpha_{k,l,t} ~\mathcal{H}_{k,l,t},
\end{equation}
where it can be assumed that the transmitted side has $N_t$ antenna elements and the received side has $N_r$, and $\mathcal{H}_{k,l,t}$ is given as
\begin{equation}
\begin{aligned}
    \mathcal{H}_{k,l,t}= \mathbf{U} \boldsymbol{a}(\theta_{k,A,l,t}) \boldsymbol{a}^H(\theta_{k,D,l,t}) \mathbf{U}^H ~~~~~~~~~~~~~~~\\
    = \frac{1}{N_t N_r} \times ~~~~~~~~~~~~~~~~~~~~~~~~~~~~~~~~~~~~~~\\
    \begin{bmatrix}
        f(\phi_{A},\phi_{D},\psi_{t,0},\psi_{r,0}) & \hdots & f(\phi_{A},\phi_{D},\psi_{t,N_t-1},\psi_{r,0})\\
        \vdots & \ddots & \vdots\\
        f(\phi_{A},\phi_{D},\psi_{t,0},\psi_{r,v}) & \ddots & f(\phi_{A},\phi_{D},\psi_{t,N_t-1},\psi_{r,v})\\
        \vdots & \ddots & \vdots\\
        f(\phi_{A},\phi_{D},\psi_{t,0},\psi_{r,N_r-1}) & \hdots & f(\phi_{A},\phi_{D},\psi_{t,N_t-1},\psi_{r,N_r-1})
    \end{bmatrix} , \\
\end{aligned}
\end{equation}
where $\phi_{A}=\frac{d}{\lambda} \sin(\theta_{k,A,l,t})$, $\phi_{D}=\frac{d}{\lambda} \sin(\theta_{k,D,l,t})$, and the analysis of $f$ is given in Appendix. A. \hfill$\blacksquare$

From the power-focusing ability of Dirichlet sinc function, it can be concluded that the power of $\mathbf{H}_{b,k,t}$ is concentrated only on a small number of elements \cite{sayeed2010continuous}. Thus, when considering the \ac{UL} transmission, the transmission of each user happens mainly through that small number of elements at the \ac{BS} side which reduces the interference between the \ac{UE}s.

\section{The Proposed Beam Tracking Algorithm} \label{sec:tracking-algorithm}
In this section, the frame transmission structure is presented first.  
Then, the evolution model is presented, while the beam tracking algorithm is proposed after that.

\subsection{Frame structure}
\begin{figure}
\begin{subfigure}{.5\textwidth}
  \centering
  \includegraphics[scale=0.28]{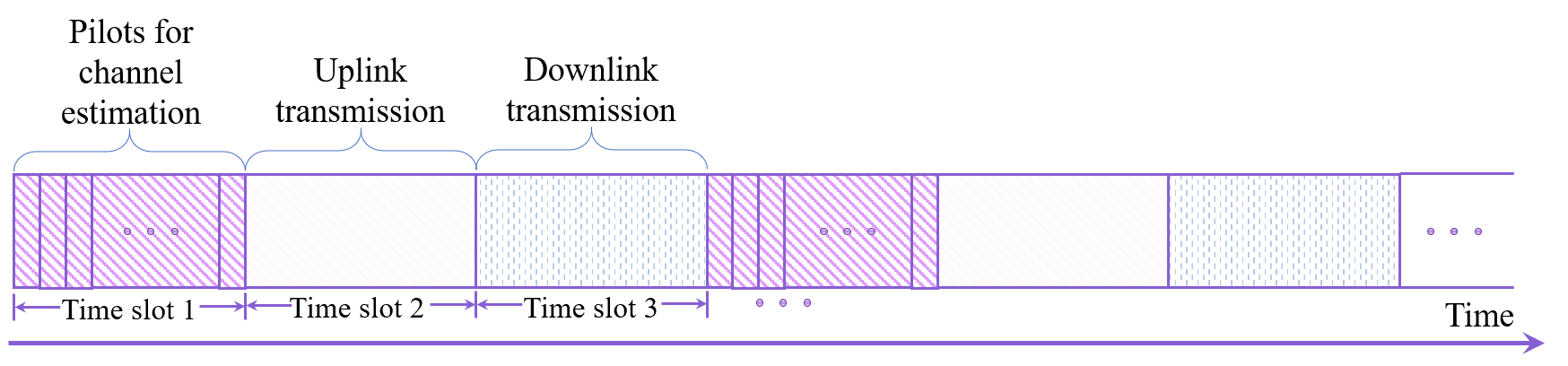}  
  \caption{}
\end{subfigure}
\begin{subfigure}{.5\textwidth}
  \centering
  \includegraphics[scale=0.29]{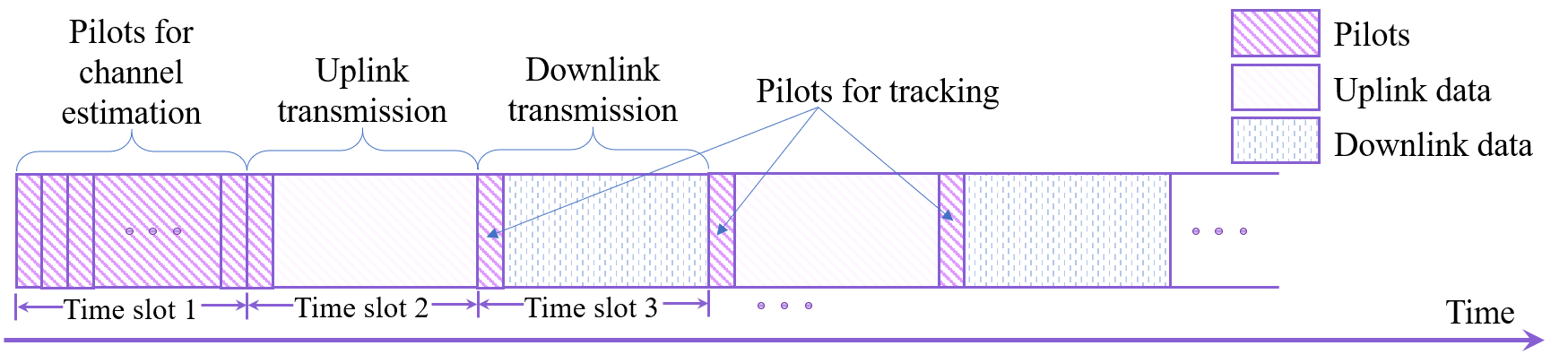}  
  \caption{}
\end{subfigure}
\caption{Frame structure for the system transmission (a) without tracking, and (b) with beam and channel tracking.}
\label{fig:frame-structure}
\end{figure}

The frame structure for the proposed beamspace \ac{MIMO} system is similar to the structure in \cite{li2017fast}, where one total slot is allocated for beamspace channel estimation. Then, assuming that all the \ac{UE}s are synchronized, in each time-slot, one pilot is allocated for beam and channel tracking in either \ac{UL} or \ac{DL} transmissions. Note that the proposed algorithm is also designed to work with only one pilot to update its parameters. The proposed frame structure for the tracking procedure is illustrated in Fig. \ref{fig:frame-structure}. 
\subsection{Evolution model}
In this work, three parameters are considered to track the beam; complex channel coefficients, \ac{AoA}, and \ac{AoD}. Therefore, in order to prepare the system for tracking, two evolution models should be presented; state and generic models.

In the state-evolution model, the state-space vector for all channel paths at time $t$ can be given as
\begin{equation} \label{equ:statespacevector}
  \boldsymbol{x}_{t} =  [\boldsymbol{\alpha}_{\Re,t} ~\boldsymbol{\alpha}_{\Im,t} ~\boldsymbol{\theta}_{D,t}~ \boldsymbol{\theta}_{A,t}]^T,
\end{equation}
where we separate the channel coefficient into real $\alpha_{\Re}$ and imaginary $\alpha_{\Im}$ parts to make sure that the angles are real along with the tracking procedures. Since Gauss–Markov model is widely adopted as a simple and effective model to characterize the fading process \cite{medard2000effect}, the evolution model for the channel coefficients can be assumed as a first-order Gauss-Markov \cite{wooldridge2016introductory,va2016beam} given as 
\begin{equation}
\label{equ:model-coefficient}
  \boldsymbol{\alpha}_{t} =  \rho \boldsymbol{\alpha}_{t-1} + \boldsymbol{\zeta}_{t-1},
\end{equation}
where $\rho$ is the channel fading correlation coefficient that characterizes the degree of time variation, and $\zeta \backsim \mathbb{C}\mathcal{N}(0, 1-\rho^2)$ \cite{jayaprakasam2017robust}.
The generic-evolution model for \ac{AoA} and \ac{AoD} follows the Gaussian process noise model \cite{zhang2016tracking, va2016beam} and given by
\begin{equation}
\label{equ:model-angles}
  \boldsymbol{\theta}_{i,t} =  \boldsymbol{\theta}_{i,t-1} + \boldsymbol{\xi}_{i,t-1},
\end{equation}
with $\xi \backsim \mathcal{N}(0,\sigma_i^2)$, $i \in \{A, D \}$.

Note that $\sigma^2$ and $(1-\rho^2)$ determines the channel variations. Higher values of $\sigma^2$ and $(1-\rho^2)$ imply fast fading channel, while lower values are used for slow fading channels. 

Based on these evolution models, the state-evolution can now be written as
\begin{equation}
\label{equ:state-evolution}
    \boldsymbol{x}_t = \mathbf{F}(\boldsymbol{x}_{t-1},\mathbf{u}_{t-1}),
\end{equation}
where $\mathbf{F}$ is a function of the generic-evolution models for complex channel coefficients, \ac{AoA}s and \ac{AoD}s as in \eqref{equ:model-coefficient} and \eqref{equ:model-angles}. $\mathbf{u} \backsim \mathcal{N}(0,\mathbf{Q})$ representing the distribution of $\zeta$ and $\xi$ with $\mathbf{Q}=diag([(1-\rho^2)/2 ~ (1-\rho^2)/2 ~ \sigma^2_A ~ \sigma^2_D])$ since the channel parameters are considered independent.

\subsection{Proposed unscented Kalman tracking filter}
In this subsection, the proposed \ac{UKF} algorithm is applied to the system models that described in Section \ref{sec:system-model} for beam and channel tracking. \ac{UKF} is considered as an advantageous to \ac{EKF} due to its ability to adapt to model changes and to overcome the weaknesses of the \ac{EKF} while having the same complexity as explained in \cite{lu2007two,montella2011kalman,thrun2002probabilistic}. This preference was proved in \cite{allotta2016unscented,allotta2016new}, where the performances of \ac{UKF} and \ac{EKF} were compared for an autonomous underwater vehicle navigation system.
Moreover, \ac{UKF} is quite suitable for a heavily nonlinear system since its estimation characteristic is not concerned by the level of nonlinearity, which makes the algorithm commonly used in many engineering fields such as integrated navigation \cite{meng2016covariance}, autonomous underwater vehicle navigation \cite{allotta2016unscented,allotta2016new}, system identification \cite{kallapur2008ukf}, target tracking \cite{zhang2013unscented}, and location tracking \cite{larew2018adaptive}.

In this paper, \ac{UKF} based algorithm is employed to track the beam. 
In order to start the tracking, a perfect channel estimation is assumed and the state space vector $\boldsymbol{x}$ is initialized from the estimated beam/channel parameters. The input of the proposed algorithm is the state space vector at time $t$ represented by its mean $\overline{\boldsymbol{x}}_t$ and covariance $\mathbf{\Sigma}_{x,t}$. The measurement/observed symbol $y_t$ is needed as an input to update the algorithm.

The state distribution for the algorithm is represented by a Gaussian random variable specified utilizing a minimal set of carefully chosen sample points. These points are called sigma points and they completely capture the true mean and covariance of $\overline{\boldsymbol{x}}_t$ and $\mathbf{\Sigma}_{x,t}$. Noting that when these points propagated through the real-time non-linear system, they can obtain the posterior mean and covariance correctly unlike the \ac{EKF} algorithm where the nonlinearity is approximated to a linear using Jacobian matrix. 
These sigma points are given as
\begin{equation}
\label{equ:sigma-points}
\begin{aligned}
    \mathbf{\chi}_{t-1}= \bigg[ \overline{\boldsymbol{x}}_{t-1} ~~~ \overline{\boldsymbol{x}}_{t-1}+ \sqrt{(m +  \Lambda  )\mathbf{\Sigma}_{x,t-1}} \\     ~~~ \overline{\boldsymbol{x}}_{t-1}- \sqrt{(m +  \Lambda  )\mathbf{\Sigma}_{x,t-1}} \bigg]  
\end{aligned}
\end{equation}
where $m$ is the number of state-space elements in $\boldsymbol{x}$ and $\Lambda$ is a scaling parameter such that $( \Lambda +m)\neq 0$ and it can control the amount sigma points spreading around the mean.
These sigma points are then propagated through the process model given in \eqref{equ:state-evolution} returning in the end a cloud of transformed points $\mathbf{\chi}_{t}$. 
The new estimated mean $\overline{\boldsymbol{x}}_t$ and covariance $\mathbf{\overline{\Sigma}}_{x,t}$ are then computed from the transformed points $\mathbf{\chi}_{t}$ as
\begin{equation}
    \overline{\boldsymbol{x}}_t = \sum_{i=0}^{2m}  \varpi^{(m)}_i ~ \mathbf{\chi}_{t,i},
\end{equation}
\begin{equation}
    \mathbf{\overline{\Sigma}}_{x,t} = \sum_{i=0}^{2m}  \varpi^{(c)}_i ~ \left(\mathbf{\chi}_{t,i}-\overline{\boldsymbol{x}}_t\right) \left(\mathbf{\chi}_{t,i}-\overline{\boldsymbol{x}}_t\right)^T + \mathbf{Q}_t,
\end{equation}
where $\varpi^{(m)}=\varpi^{(c)}=\frac{1}{2(m+ \Lambda)}$ for $i=1,2,\cdots,2m$ while $\varpi^{(m)}_0=\frac{\Lambda}{\Lambda +m}$ and $\varpi^{(c)}_0=\frac{\Lambda}{\Lambda +m}+(1-\gamma^2+\beta)$, noting that the weights are normalized to satisfy $\sum_{i=0}^{2m}\varpi_i=1$.  
$\beta$ is used to incorporate prior knowledge of the distribution of $\boldsymbol{x}$, which is set to 2 as an optimum value for a Gaussian distribution \cite{wan2000unscented}, while $\gamma \in [0, 1]$ is a scaling parameter used to identify $\Lambda$ given that $\Lambda = \gamma ^2~(m+ \kappa )-m$ where $\gamma$ and $\kappa$ are scaling parameters that are responsible of determining the spreading of sigma points around the mean $\overline{\boldsymbol{x}}$.

The transformed sigma points $\mathbf{\chi}_{t}$ are then propagated through the nonlinear observation model $g(\mathbf{\chi}_{t})$. Considering the system is tracked in the \ac{DL} mode, based on \eqref{equ:received-signal-downlink}, the observation function in a beamspace domain is given as
\begin{equation}
\label{equ:observation-DL}
   \mathbf{Z}^{({\operatorname{DL}})}_t = \mathbf{g}_{\operatorname{DL}}(\mathbf{\chi}_{t}) = \boldsymbol{w}^H_k \mathbf{H}_{b,k,t}(\mathbf{\chi}_{t}) \mathbf{p}_k, 
\end{equation}
where ${s}_k$ is considered as a unit pilot symbols. 

For the \ac{UL} beamspace transmission, the \ac{BS} is responsible of tracking multi-beam simultaneously. Therefore, referring to \eqref{equ:received-signal-uplink}, the observation function will lead to multi measurements and is given as
\begin{equation}
\label{equ:observation-UL}
   \mathbf{Z}^{({\operatorname{UL}})}_t = \mathbf{g}_{\operatorname{UL}}(\mathbf{\chi}_{t}) = \boldsymbol{W}^H \mathbf{H}_{b,t}(\mathbf{\chi}_{t}) \mathbf{P} \mathbf{D} . 
\end{equation}
This class of filter can be relevant even when there is a disconnectedness in nonlinear functions $\mathbf{F}$ and $\mathbf{g}$.

The mean $\overline{\mathbf{z}}_t$ and covariance $\mathbf{\Sigma}_{z,t}$ of the transformed observations are measured as
\begin{equation}
    \overline{\mathbf{z}}_t = \sum_{i=0}^{2m}  \varpi^{(m)}_i ~ \mathbf{Z}_{t,i},
\end{equation}
\begin{equation}
    \mathbf{\Sigma}_{z,t} = \sum_{i=0}^{2m}  \varpi^{(c)}_i ~ \left(\mathbf{Z}_{t,i}-\overline{\mathbf{z}}_t\right) \left(\mathbf{Z}_{t,i}-\overline{\mathbf{z}}_t\right)^T + \mathbf{\sigma}_v^2,
\end{equation}

After that, in order to measure the filter gain $K_t$, the cross covariance between the transformed observation and the transformed sigma points is needed and can be expressed as follow
\begin{equation}
    \mathbf{\Sigma}_{xz,t} = \sum_{i=0}^{2m}  \varpi^{(c)}_i ~ \left(\mathbf{\chi}_{t,i}-\overline{\mathbf{x}}_t\right) \left(\mathbf{Z}_{t,i}-\overline{\mathbf{z}}_t\right)^T .
\end{equation}
Therefore, the gain is defined as
\begin{equation}
\label{equ:filter-gain}
    K_t = \mathbf{\Sigma}_{xz,t}  \mathbf{\Sigma}_{z,t}^{-1}.
\end{equation}
Lastly, the posterior state-space vector and its covariance are updated as
\begin{equation}
\label{equ:state-space-update}
    \mathbf{x}_t = \overline{\mathbf{x}}_t + K_t (\mathbf{y}_t - \overline{\mathbf{z}}_t),
\end{equation}
\begin{equation}
\label{equ:covariance-update}
    \mathbf{\Sigma}_{x,t} = \overline{\mathbf{\Sigma}}_{x,t} + K_t \mathbf{\Sigma}_{z,t} K_t^T.
\end{equation}
The tracking will be repeated for each measurement update on $t$-th time index until the algorithm fails to track due to extreme changes on the channel and new channel parameter estimation is performed.

It should be noted that the algorithm addresses the approximation issues of the \ac{EKF} by using \ac{UT}. This concept is generated under the fact that the approximation of a given distribution, by using a fixed number of parameters, is easier than approximating an arbitrary nonlinear function \cite{gordon2004beyond}. Following this approach, the \ac{UT} obtains a set of $2m+1$ sigma points, deterministically chosen as presented in \eqref{equ:sigma-points}. The uniqueness of the \ac{UKF} algorithm is in the way of selecting these points: numbers, values, and weights. The sigma point method results in a more accurate computation of the nonlinear system tracking where the accuracy is increased as the set of sigma points increases. However, the amount to pay is a significant increase in computational cost as $2m+1$ parameters are additionally performed in the system.
Therefore, since the spreading of the sigma points can control the accuracy of the \ac{UKF} algorithm, in this work, the spreading of these sigma points are controlled so that the modified \ac{UKF} algorithm provides better performance during the tracking time without additional performed parameters.   

\begin{figure}
\centering
\includegraphics[scale=0.26]{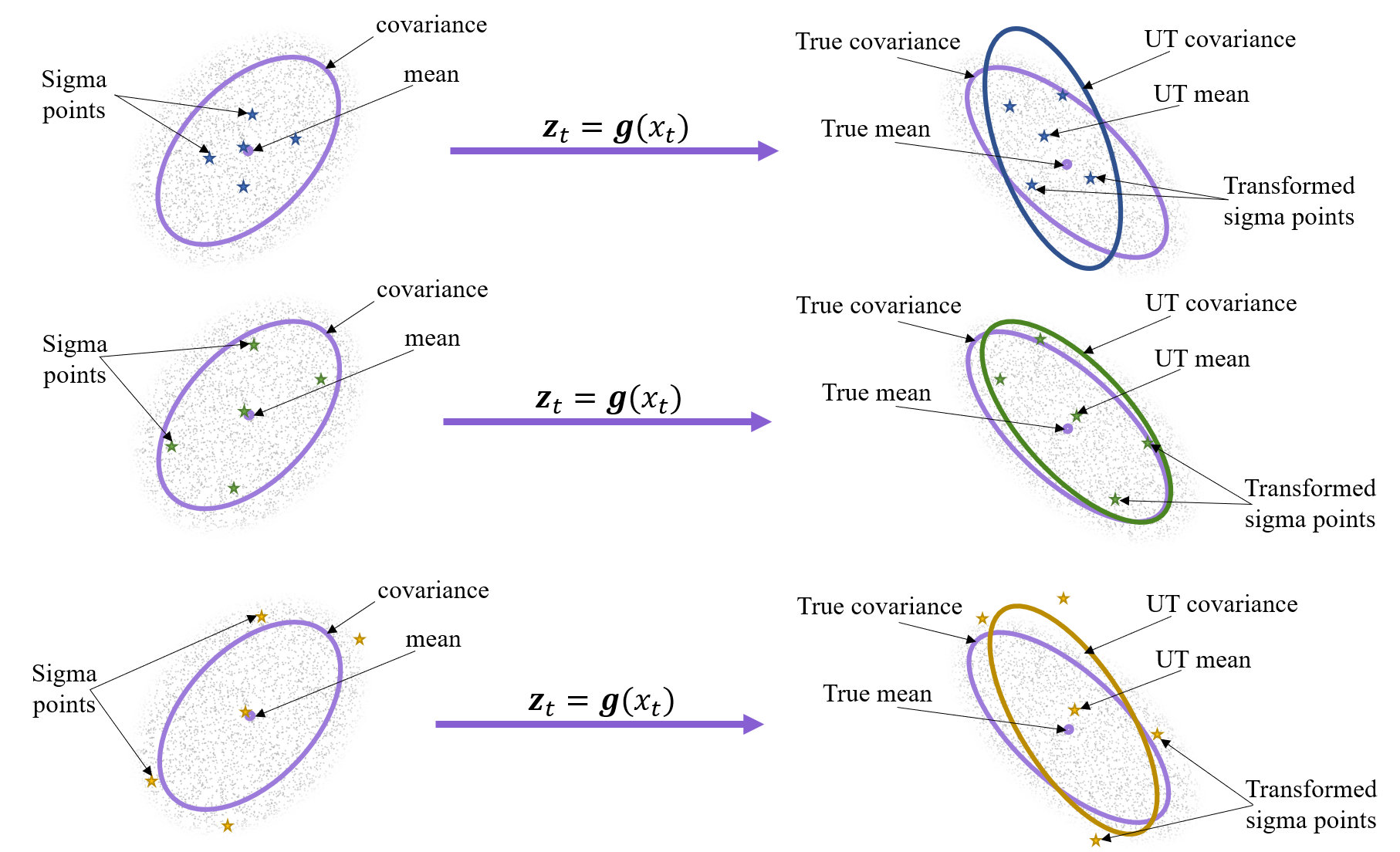}
\caption{Effect of sigma points spreading on the transformed procedure in \ac{UKF} algorithm.}
\label{fig:sigma-points}
\end{figure}

The effect of different sigma points spreading around the true mean is illustrates in Fig. \ref{fig:sigma-points}, where it is clearly shown that choosing different spreading can lead to different mean and covariance than the true ones. 

In order to optimize the spreading of the sigma points, optimal values for the scaling parameters $\gamma$ and $\kappa$ are chosen so that the innovation in \eqref{equ:state-space-update} is minimized, and it is formulated as 
\begin{equation}
\label{equ:optimization}
    \{ \gamma_t^+, \kappa_t^+ \} = \underset{\gamma_t, \kappa_t}{\min} \{ \mathbf{y}_t - \overline{\mathbf{z}}_t \}.
\end{equation}

The proposed \ac{UKF} tracking algorithm scheme is summarized in Algorithm \ref{algorithm:tracking}.
\begin{algorithm}
\DontPrintSemicolon
  \KwInput{Received noisy signal $\mathbf{y}_t$ in \eqref{equ:received-signal-uplink} for \ac{UL} or in \eqref{equ:received-signal-downlink} for \ac{DL}, initial state-space vector $\boldsymbol{x}_0$ }
  \KwOutput{$\boldsymbol{x}_t$, $\mathbf{\Sigma}_{x,t}$}
  \textbf{Estimate} the total beamspace channel $\mathbf{H}_{b}$.\;
  \For{t = 1,2,3,$\dots$}
  { 
    \textbf{Optimize} the $\gamma_t$ and $\kappa_t$ using \eqref{equ:optimization} for the first time slot only. \;
    \textbf{Calculate} $\mathbf{\chi}_{t-1}$ as in \eqref{equ:sigma-points}.\;
    \textbf{Update} $\mathbf{\chi}_{t-1}$ to $\mathbf{\chi}_{t}$ using \eqref{equ:state-evolution}.\; 
    \textbf{Propagate} $\mathbf{\chi}_{t}$ through $\mathbf{g}(\mathbf{\chi}_{t})$ using \eqref{equ:observation-DL} for \ac{DL} transmission or \eqref{equ:observation-UL} for \ac{UL} transmission. \;
    \textbf{Calculate} the filter gain $K_t$ using \eqref{equ:filter-gain}.\;
    \textbf{Update} the state-space $\boldsymbol{x}_t$ and its covariance $\mathbf{\Sigma}_{x,t}$ using \eqref{equ:state-space-update} and \eqref{equ:covariance-update}.\;
    \textbf{Return} $\boldsymbol{x}_t$, $\mathbf{\Sigma}_{x,t}$
  }
\caption{The proposed channel tracking.}
\label{algorithm:tracking}
\end{algorithm}
\begin{figure*}
\begin{subfigure}{.5\textwidth}
  \centering
  \includegraphics[scale=0.6]{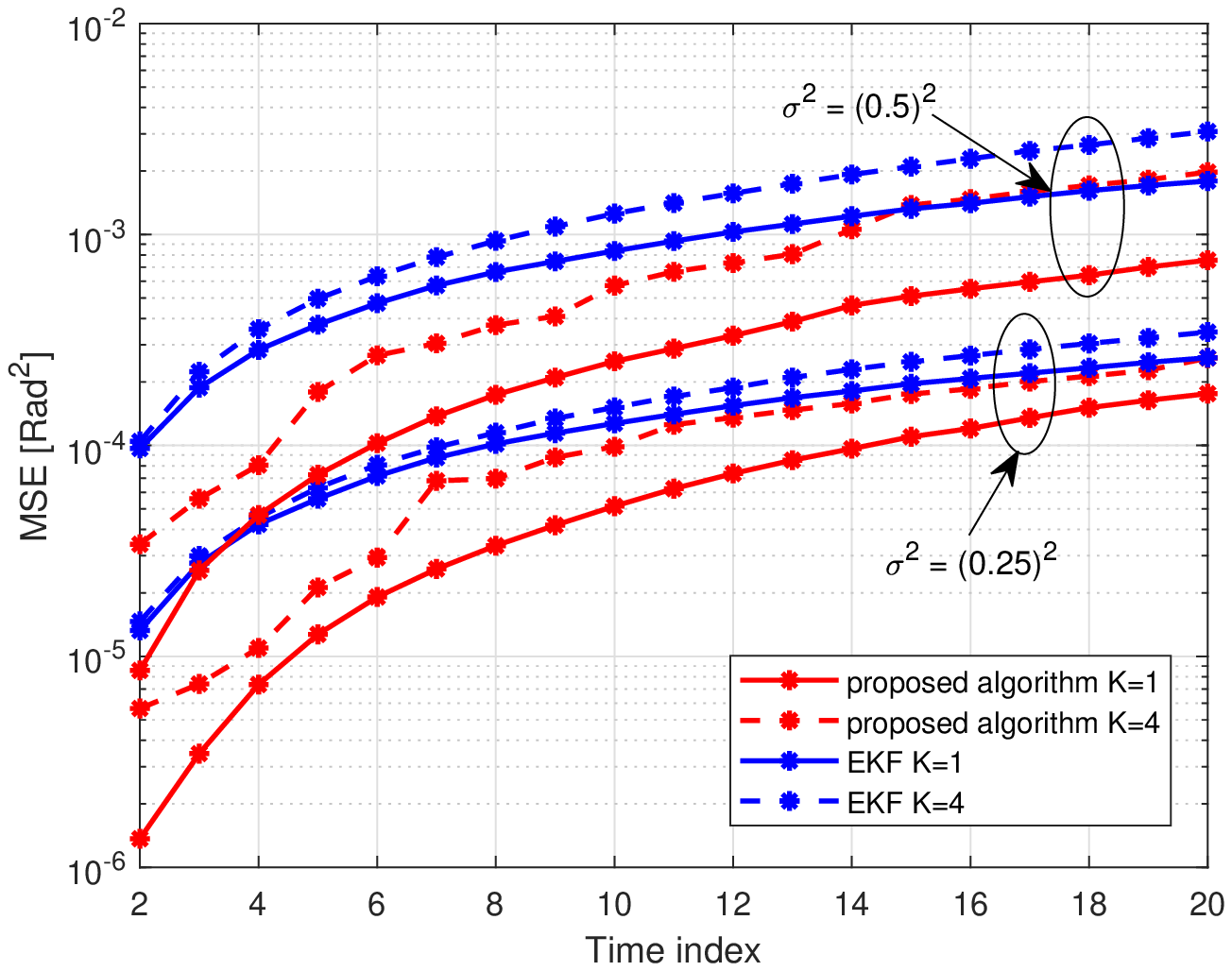}  
  \caption{}
\end{subfigure}
\begin{subfigure}{.5\textwidth}
  \centering
  \includegraphics[scale=0.6]{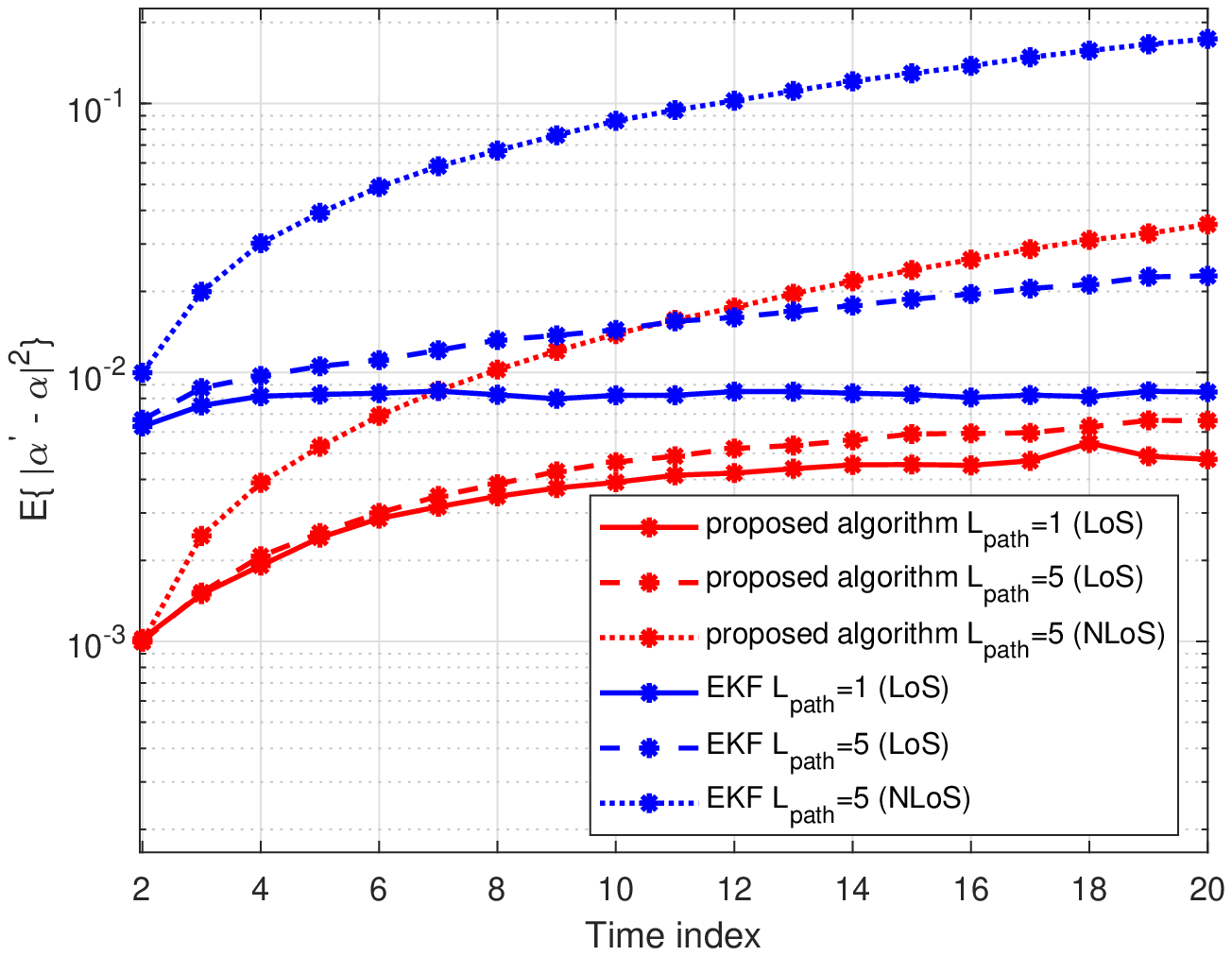}  
  \caption{}
\end{subfigure}
\caption{The \ac{MSE} of (a) \ac{AoA} for the single- and multi-user system, and (b) the channel tracking for a single-user system at the \ac{DL} transmission.}
\label{fig:error-downlink}
\end{figure*}

\section{Numerical Results} \label{sec:numerical-results}
In this section, the numerical results are presented to examine the performance of the proposed tracking algorithm in several scenarios. We explore the impact of different number of \ac{UE}s in the system and provide a comparison with \ac{EKF} based tracking algorithm \cite{zhang2016tracking,va2016beam}. The performance of each tracking method is shown by calculating the \ac{MSE} for the angles which is given as
\begin{equation}
    J_k=E\left[ |\mathbf{\widehat{\theta}}_{t}-\mathbf{\theta}_{t}|^2_2 \right],
\end{equation}
where $\mathbf{\widehat{\theta}}_t$ is the tracked angles while $\mathbf{\theta}_{t}$ is the estimated angles considering no tracking is hired in the system. The channel fading correlation $\rho$ for the system is set to be $\rho=0.99$ corresponds to slow fading in channel coefficient. A random initialization for the tracking parameters $\theta_A$ and $\theta_D$ is given from a uniform distribution $\mathcal{U}(0,\pi)$, and $\alpha$ as a complex standard normal distribution. Due to the similarity in performance for \ac{AoA} and \ac{AoD}, only the \ac{AoA} performance is given. It is assumed that the number of \ac{RF} chains at \ac{BS} side is equal to the number of served \ac{UE}s in both \ac{DL} and \ac{UL} transmissions ($N_{\operatorname{RF}}=K$).
Since a reliable high-data transmission in mmWave can be provided with only a few number of path components \cite{han2015large}, the \ac{DL} and \ac{UL} transmissions are assumed to have one path component between each user and the \ac{BS} during the analysis. As well, based on the sparse nature of the beamspace channel at mmWave frequencies, we can select only a small number of dominant beams to reduce the effective dimension of \ac{MIMO} system without obvious performance loss. So, few \ac{UE}s are served in the system. Each \ac{MSE} plot is obtained by averaging over 10000 simulation runs.

\subsection{Beam and channel tracking in the \ac{DL} transmission}
In this subsection, the proposed algorithm is used for tracking at the user side, where the \ac{BS} can serve multiple \ac{UE}s. The system parameters setup is given in Table \ref{table:simulationDL}.

\begin{table}
\caption{Simulation configuration in the \ac{DL} transmission. }
\centering
\begin{tabular}{l|l}
\hline
\multicolumn{1}{c|}{Parameters}                          &      \multicolumn{1}{c}{Value}   \\ \hline
Operating frequency $f_c $                               &  28 GHz         \\ \hline
Channel paths for the tracked user  $L_k$                &  1, 5             \\ \hline
Channel paths for other \ac{UE}s $l$                     &  1             \\ \hline
Antenna array elements at \ac{BS} $N_{\operatorname{BS}}$ &  16              \\ \hline
Antenna array elements at all \ac{UE}s $N_k$             &  8              \\ \hline
Distance between antenna elements $d$                    &  $\lambda/2$      \\ \hline
Angle speed variation $\sigma^2_A=\sigma_D^2=\sigma^2$   &  $(0.25)^2$, $(0.5)^2$    \\ \hline
Tracking duration                                        &  20 time slot      \\ \hline
Channel fading correlation coefficient $\rho$            & 0.99     \\ \hline
\ac{SNR}                                                 & 20 dB     \\ \hline
Number of \ac{UE}s in the system $K$                     & 1, 4      \\ \hline
\end{tabular}
\label{table:simulationDL}
\end{table}

Fig. \ref{fig:error-downlink}a depicts the \ac{MSE} beam tracking performance for single- and multi-user system in the \ac{DL} transmission at different speed variation angles. It is seen that as the variation $\sigma^2$ increases from $0.25^2$ to $0.5^2$, the measured error of the proposed tracking algorithm increases from $10^{-4}$ to $5\texttt{x}10^{-3}$ for single user scenario. Although the proposed algorithm performs poorly when the number of \ac{UE}s increases in the system, it can beat the performance of the conventional \ac{EKF} algorithm. It is clearly shown the proposed algorithm outperforms the conventional \ac{EKF} up to 85\% performance enhancement in a high speed variation system. 

Fig. \ref{fig:error-downlink}b illustrates the channel tracking performance for both the proposed and the conventional \ac{EKF} algorithms for a single-user system $K=1$ at angle speed variation of $\sigma^2=(0.25)^2$. It is noticed that the proposed algorithm gives up to 44\% performance enhancement in tracking single path, while it reaches around 66\% enhancement for tracking the \ac{LoS} path and more than 80\% for tracking the \ac{NLoS} paths in a user that has 5 resolvable channel paths.

\begin{figure*}
\begin{subfigure}{.5\textwidth}
  \centering
  \includegraphics[scale=0.6]{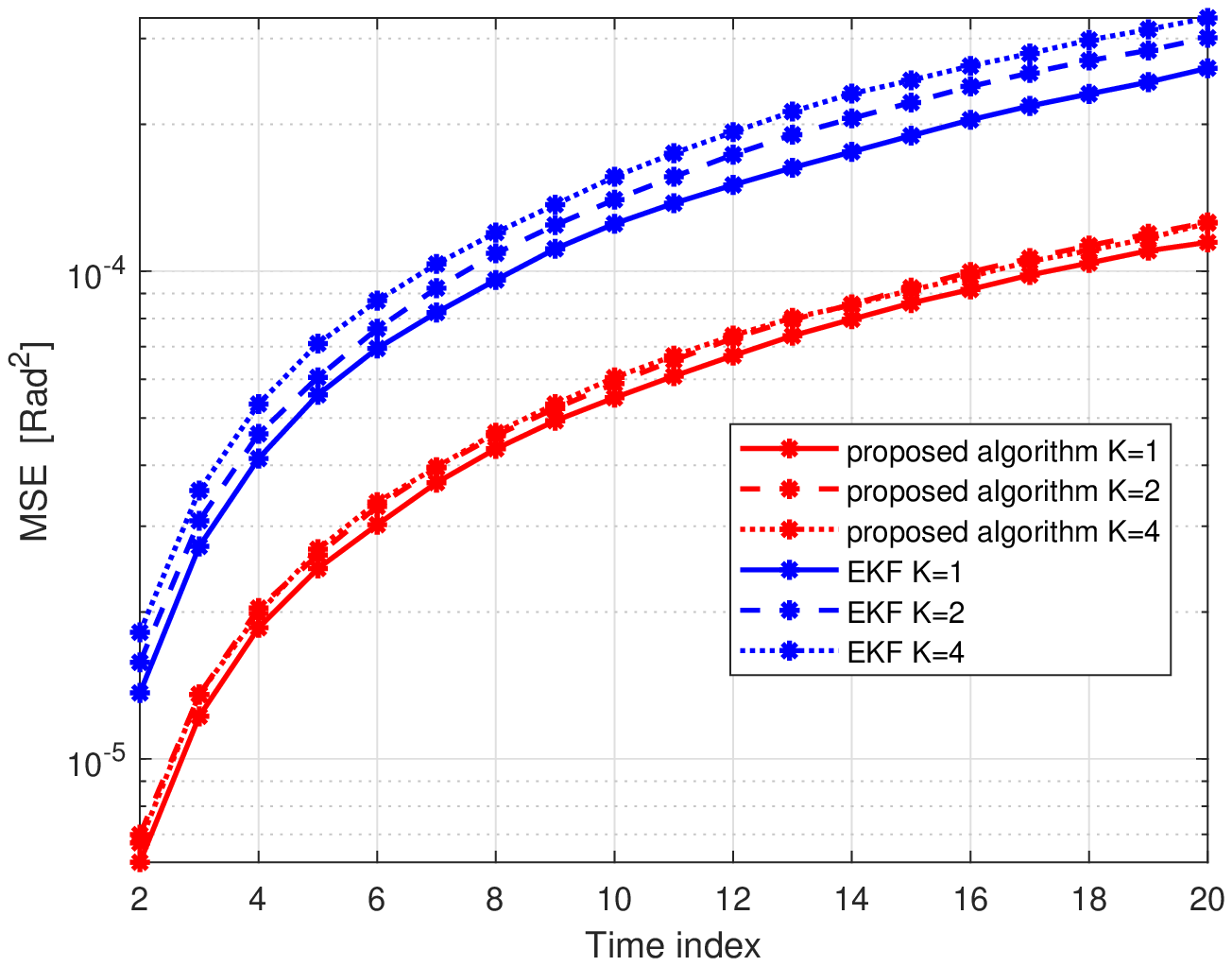}  
  \caption{}
\end{subfigure}
\begin{subfigure}{.5\textwidth}
  \centering
  \includegraphics[scale=0.6]{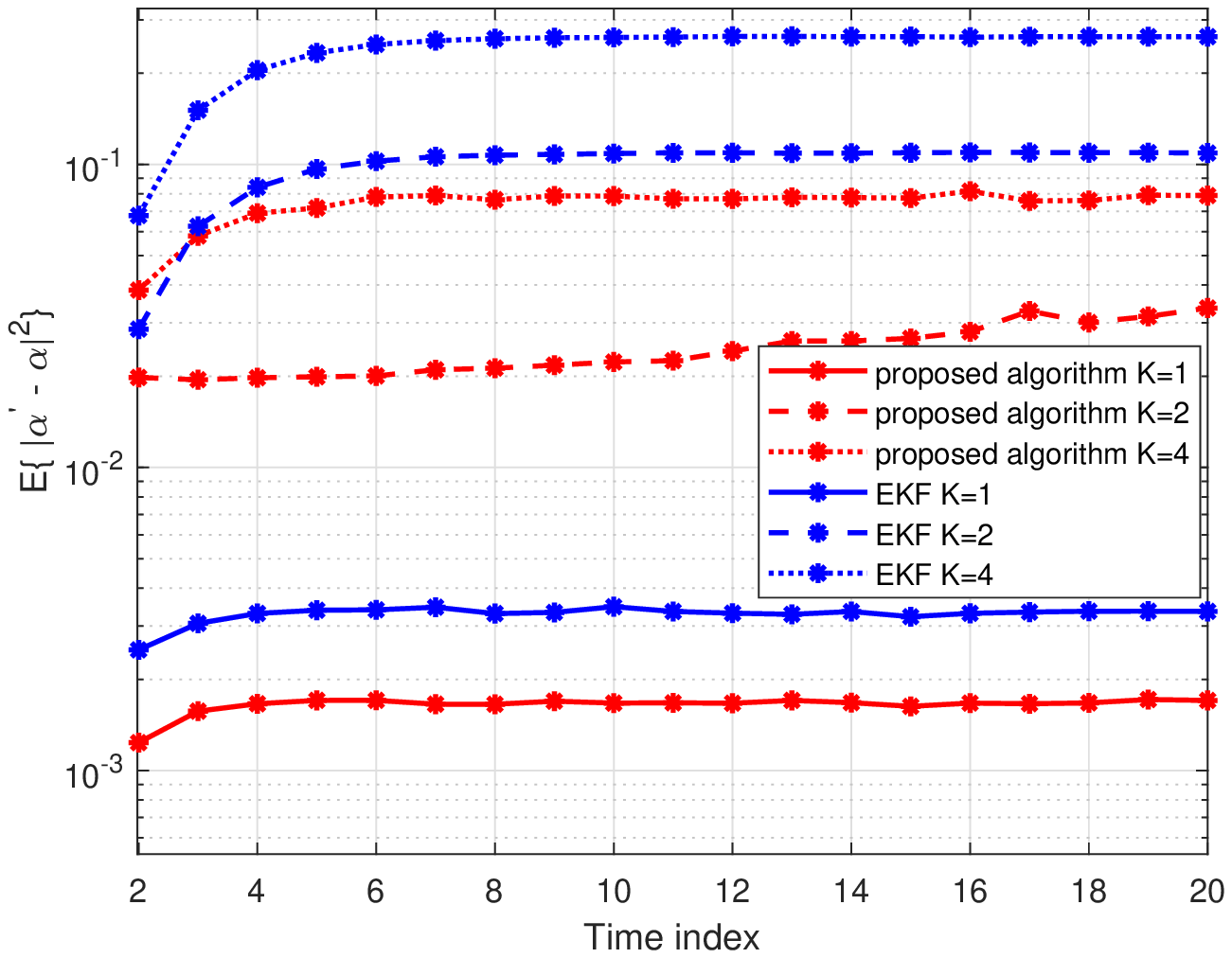}  
  \caption{}
\end{subfigure}
\caption{The \ac{MSE} of (a) \ac{AoA} for different number of \ac{UE}s, and (b) its channel tracking  at the \ac{UL} transmission.}
\label{fig:error-uplink}
\end{figure*}

\subsection{Beam and channel tracking in the \ac{UL} transmission}
In this subsection, the \ac{UL} transmission system is provided where the \ac{UE}s are assumed to transmit at same power level and received at the same time by the \ac{BS}. Note that same power transmission by the \ac{UE}s is the worst case scenario due to the high level of inter-user interference observed at the receiver. The proposed algorithm is optimized its parameters according to receiving the \ac{UE}s' signals together. So that all \ac{UE}s have been tracked jointly and simultaneously. However, the \ac{EKF} algorithm in this case is employed to track each user channel separately in the presence of receiving signal from other \ac{UE}s. The system parameters setup for the \ac{UL} transmission are given in Table \ref{table:simulationUL}. 

\begin{table}
\caption{Simulation configuration in the \ac{UL} transmission.}
\centering
\begin{tabular}{l|l}
\hline
\multicolumn{1}{c|}{Parameters}                          &      \multicolumn{1}{c}{Value}   \\ \hline
Operating frequency $f_c $                               &  28 GHz         \\ \hline
Channel paths between each \ac{UE} and the \ac{BS}  $L_k$ &  1             \\ \hline
Antenna array elements at \ac{BS} $N_{\operatorname{BS}}$ &  16              \\ \hline
Antenna array elements at all \ac{UE}s $N_k$             &  8              \\ \hline
Distance between antenna elements $d$                    &  $\lambda/2$      \\ \hline
Angle speed variation for all \ac{UE}s $\sigma^2_A=\sigma_D^2=\sigma^2$   &  $(0.35)^2$     \\ \hline
Tracking duration                                        &  20 time slot      \\ \hline
channel fading correlation coefficient  $\rho$            & 0.99     \\ \hline
Averaged \ac{SNR} from each \ac{UE}                         & 0 dB     \\ \hline
Number of \ac{UE}s in the system $K$                     & 1, 2, 4      \\ \hline
\end{tabular}
\label{table:simulationUL}
\end{table}

Fig. \ref{fig:error-uplink}a demonstrates the \ac{MSE} performance of \ac{AoA} beam tracking for the single- and multi-user beamspace \ac{MIMO} systems while Fig. \ref{fig:error-uplink}b illustrates the channel tracking performance for the same system at the \ac{UL} transmission. According to Fig. \ref{fig:error-uplink}a, the effectiveness of the proposed algorithm in tracking multiple beams jointly is clearly shown with 62\% performance enhancement compared to the conventional \ac{EKF} method. As well, the degradation in performance between different number of \ac{UE}s in the system is negligible. However, as shown in Fig. \ref{fig:error-uplink}b, the performance gap between the proposed algorithm in a single-user system and multi-user system is notable where there is around 90\% reduction in performance between one-user system and two-user system. Noting that the gap between the proposed algorithm and the conventional \ac{EKF} based method is increased as well.

\section{Conclusion} \label{sec:conclusion}
A novel multi-beam joint-tracking algorithm based on \ac{UKF} filter was designed for multi-user beamspace \ac{MIMO} systems using \ac{LAA}. The proposed algorithm is employed to track the channel parameters; \ac{AoA}, \ac{AoD}, and channel coefficient, of multi-beam jointly. The algorithm avoids Jacobian and/or Hessian matrices computation to provide a linear estimator without any approximation as it is the case in the conventional \ac{EKF} based method. Two implementations were investigated for the beam and channel tracking using the proposed algorithm: 1) single-beam tracking at the \ac{UE} side in the \ac{DL} transmission in the presence of other \ac{UE}s' beams interference, and 2) multi-beam joint-tracking at the \ac{BS} side in the \ac{UL} transmission system. 
Note that the proposed algorithm optimized the sigma points spreading parameters of \ac{UKF} method which enables us to efficiently track multiple \ac{UE}s simultaneously. This leads to enhancing the tracking performance by reducing overall interference in the \ac{UL} transmission.
The numerical results showed that the proposed algorithm can provide up to 85\% performance enhancement in tracking performance compared to conventional \ac{EKF} based method in high mobility systems. The proposed algorithm can be implemented in a highly mobile environment to enhance mobility support by detecting changes in the channel for high-speed users. Also, the proposed algorithm is feasible for devices with limited computation and processing capabilities. As a future work, performance analysis of the presented works can be utilized in a laboratory environment.

\section*{Acknowledgement}
The work of H. Arslan was supported by the Scientific and Technological Research Council of Turkey under Grant No. 116E078.

\begin{appendices}

\section{Analysis of $f$}
Given that $\phi_{A}=\frac{d}{\lambda} \sin(\theta_{k,A,l,t})$, $\phi_{D}=\frac{d}{\lambda} \sin(\theta_{k,D,l,t})$, and $\mathbf{U} \boldsymbol{a}(\theta_{k,A,l,t}) \boldsymbol{a}^H(\theta_{k,D,l,t}) \mathbf{U}^H$ is equal to
\begin{equation}
\begin{aligned}
    \mathcal{H}_{k,l,t} =
        \begin{bmatrix}
            \mathbf{u}^H(\psi_{r,0}) \\
            \mathbf{u}^H(\psi_{r,1}) \\
            \hdots \\
            \mathbf{u}^H(\psi_{r,N_r-1}) 
        \end{bmatrix}
     \mathbf{u}(\phi_A) \times ~~~~~~~~~~~~~~~~~~\\ \mathbf{u}^H(\phi_D)
        \begin{bmatrix}
            \mathbf{u}(\psi_{t,0}) & \mathbf{u}(\psi_{t,1}) & \cdots & \mathbf{u}(\psi_{t,N_t-1}) 
        \end{bmatrix}
\end{aligned}
\end{equation}

Therefore, each element in the matrix can be given as
\begin{equation}
\label{equ:H-element}
    \left[\mathcal{H}_{k,l,t}\right]_{v,c} = \sum_{q} \sum_{i} e^{-j2\pi \left[ q i \phi_A - q \phi_D + q \psi_{t,c} - i \psi_{r,v} \right]} ,
\end{equation}
where $q=\{ \frac{-(N_t-1)}{2},\cdots, \frac{(N_t-1)}{2} \}$ and $i=\{ \frac{-(N_r-1)}{2},\cdots, \frac{(N_r-1)}{2} \}$.

Simplifying \eqref{equ:H-element} as
\begin{equation}
\begin{aligned}
    \left[\mathcal{H}_{k,l,t}\right]_{v,c} = \sum_{q} e^{-j2\pi \left[\psi_{t,c}-\phi_D\right]q} \sum_{i} e^{-j2\pi \left[ q \phi_A  -  \psi_{r,v} \right]i} , ~~\\
    = \sum_{q} e^{-j2\pi \left[\psi_{t,c}-\phi_D\right]q} \frac{\sin (\pi N_r ( q \phi_A  -  \psi_{r,v})}{\sin (\pi ( q \phi_A  -  \psi_{r,v}))} , \\
    = \sum_{q} e^{-j2\pi \left[\psi_{t,c}-\phi_D\right]q} f_{N_r}(q \phi_A  -  \psi_{r,v}), ~~~~~~\\
    = f\left( \phi_A, \phi_D, \psi_{t,c}, \psi_{r,v} \right), ~~~~~~~~~~~~~~~~~~~~~~~
\end{aligned}
\end{equation}
where $f_N(\varphi)$ is the Dirichlet sinc function with a maximum of $N$ at $\varphi=0$. It is noticed that each element in $\mathcal{H}_{k,l,t}$ is a summation of Dirichlet function with a phase shift of $\psi_{t,c}-\phi_D$ which has the same capabilities of the original Dirichlet sinc function. Therefore, the power in $\mathcal{H}_{k,l,t}$ is concentrated only on a small number of elements due to the power-focusing ability of $f_N(\varphi)$ \cite{sayeed2010continuous}.

\end{appendices}

\balance


\vfill

\end{document}